\documentclass[11pt]{article}
\usepackage[margin=1in]{geometry}
\usepackage{graphicx}

\def\beq{\begin{equation}}
\def\eeq#1{\label{#1}\end{equation}}
\def\eeqn{\end{equation}}

\def\beqa{\begin{eqnarray}}
\def\eeqa#1{\label{#1}\end{eqnarray}}
\def\eeqan{\end{eqnarray}}

\let\bar=\overbar

\def\Dslash{\not{\hbox{\kern-4pt $D$}}}
\def\dslash{\not{\hbox{\kern-2pt $\del$}}}

\def\msb{{\bar{\ssstyle M \kern -1pt S}}}

\def\Title#1{\begin{center} {\Large {\bf #1} } \end{center}}
\def\Author#1{\begin{center} {\normalsize {\sc #1} } \end{center}}
\def\Institution#1{\begin{center} {\normalsize {\it #1} } \end{center}}
\def\Abstract#1{\noindent {\normalsize {\bf Abstract:} {\normalfont #1}}}
\def\Conference{\vspace{4mm}\begin{raggedright} {\normalsize {\it Poster presented at the 2019 Meeting of the Division of Particles and Fields of the American Physical Society (DPF2019), July 29--August 2, 2019, Northeastern University, Boston, C1907293.} } \end{raggedright}\vspace{4mm}}

\begin{document}

%
%

\Title{Enhancing the LBNF Physics Programs With A high-resolution Magnetized Detector}

\Author{Hongyue Duyang, Bing Guo, Sanjib R. Mishra, Roberto Petti}

\Institution{Department of Physics and Astronomy\\ University of South Carolina, Colombia, South Carolina, United States}

\Abstract{The Long-Baseline Neutrino Facility (LBNF) offers an unprecedented intensity for neutrino physics. We discuss a proposal to enhance the LBNF physics programs at Fermilab with the addition of a high-resolution magnetized detector at the near site. The detector is largely based upon reusing an existing superconducting magnet and electromagnetic calorimeter, supplemented with a new low-density straw tube tracker. We also discuss the physics opportunities of the proposed design for the long-baseline oscillation analysis, as well as for precision measurements of (anti)neutrino interactions.}

\Conference

%
%

\section{Introduction}
The proposed Long-Baseline Neutrino Facility (LBNF) offers an unprecedented intensive neutrino beam for rich physics topics, including the measurement of neutrino oscillations at the Deep Underground Neutrino Experiment (DUNE) \cite{dune}.  
Precise knowledge of the neutrino beam, including its stability, total number of neutrinos and the neutrino energy spectrum, is critical to the success of LBNF physics programs. 
We propose to build a low-density, high-resolution, magnetized detector at the near site of the LBNF beamline 
to provide monitoring of the neutrino beam and precise measurement of the neutrino flux.   
The detector also offers constraint of nuclear smearing and neutrino energy scale for the DUNE experiment. 
It could be part of the DUNE near detector complex, or an stand-alone project by itself. 
The detector is largely based upon reusing an existing superconducting magnet and electromagnetic calorimeter, supplemented with a new low-density straw tube tracker (STT).

\section{The Straw-Tube Tracker}

\begin{figure}[htb]
\centering
\includegraphics[height=4.0in]{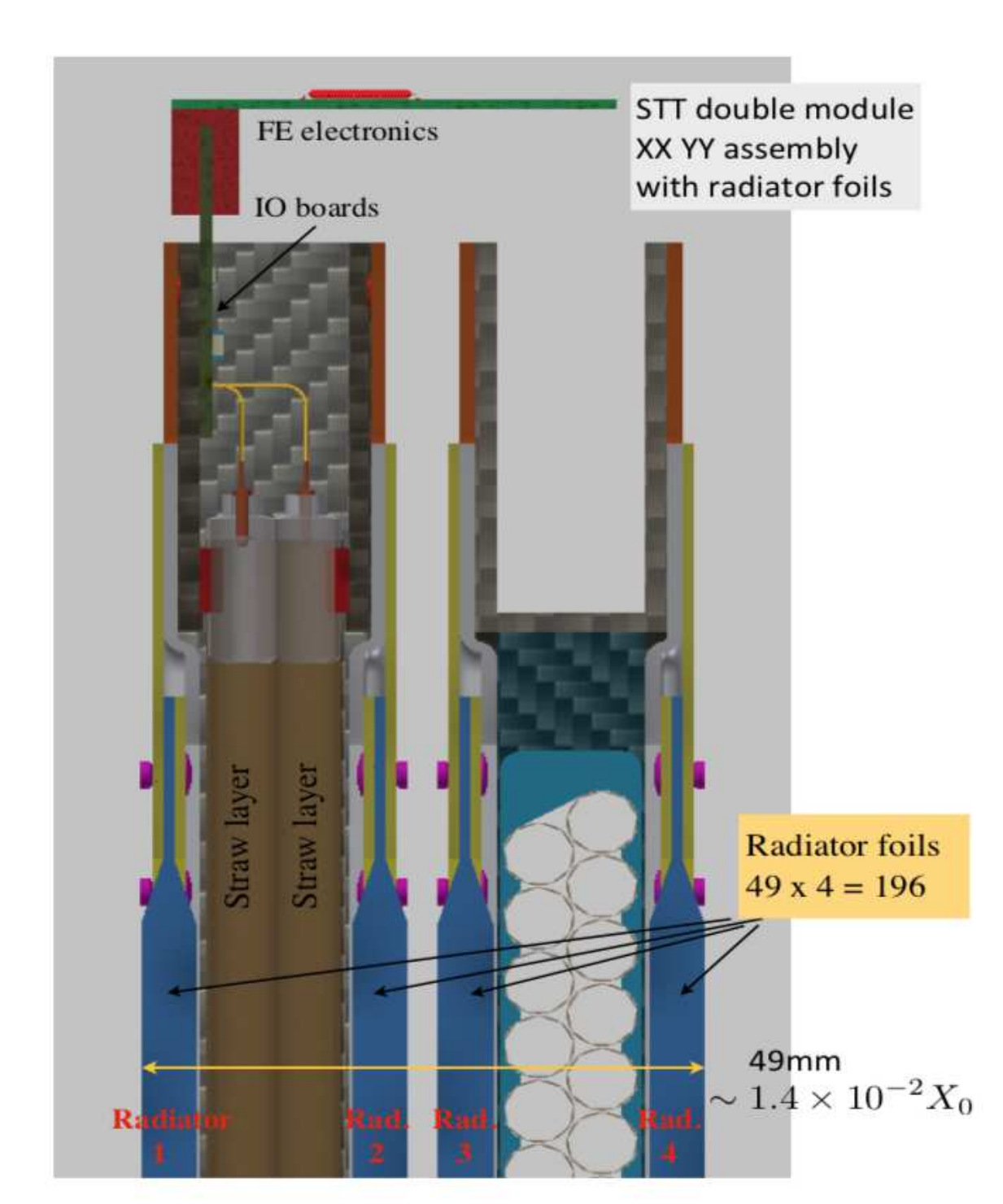}
\caption{Left: Drawing of one example STT module with straw-tubes in XXYY directions and four thin polypropylene CH$_2$ targets. }
\label{fig:stt_module}
\end{figure}

The central part of the proposed detector is a straw-tube tracker, inserted into the magnetic volume inside the magnet.
It consists of modules of low-mass straw-tubes alternated with thin layers of various target materials (Fig. \ref{fig:stt_module}).  
Each module includes four layers of straw-tubes with 5 mm diameter in orthogonal directions (XXYY), mylar walls with total thickness 20 $\mu$m and 1000 Angstroms Al coating, and tungsten wire with 20 $\mu$m diameter.
The readout of the straws can be the ASIC chip VMM3 developed at BNL and CERN for the muon chambers of LHC experiments which provides both fast timing and energy deposition.
The mass targets provide 95\% of the total detector mass, and the straw-tubes account for the rest 5\%.
The target materials include 5\,tons of CH$_2$ radiators and 600 kg of graphite target, optimized for the measurement of neutrino-hydrogen interactions, which is discussed in section 4.
The radiator target consists of 49 15\,$\mu$m thick polypropylene (C$_3$H$_6$)$_n$ foils, interspaced by 120 \,$\mu$m air gaps. 
The average density of the straw-tube tracker is 0.16 g/cm$^3$, and the radiation length $X_{0}\sim$3.5\,m. 
The tracking sampling is about 0.15 (0.36)\%$X_0\perp$($\parallel$). 
The targets can be further fine-tuned to achieve certain fiducial mass and resolutions. 
Other nuclear targets such as Ar, Ca, Fe etc., are also possible for the study of $A$-dependence of neutrino-nucleus interactions.
In particular, Ar target can be inserted as either drift tubes (2-in inner diameter with 0.04-in thick wall) of high-pressure Ar gas, or an active liquid Ar target located upstream and followed by two STT double modules without radiators. 
The total tracking volume is 3.5\,m$\times$3.5\,m$\times$4\,m = 49\,m$^3$.
The STT provides excellent momentum ($\sim$3\%) and angular ($\sim$2\,mrad) resolutions, as well as particle identification by both ionization signal dE/dx and the transition radiation produced by e$^{\pm}$ in the radiator foils. 



\section{The Magnet and Calorimeter}

\begin{figure}[htb]
\centering
\includegraphics[height=3.0in]{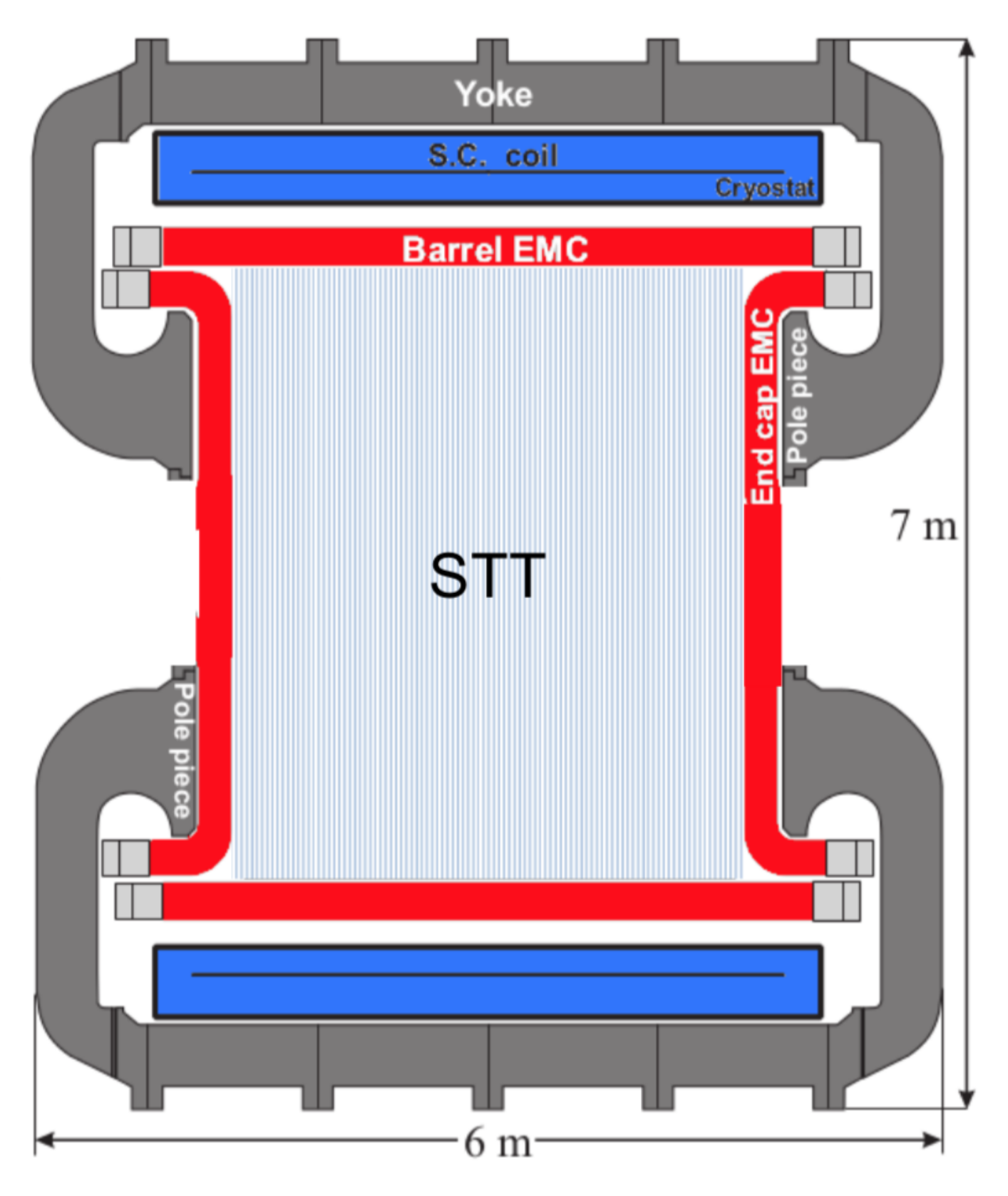}
\caption{Drawing of the top view of the STT with KLOE magnet and calorimeter. }
\label{fig:kloe-stt}
\end{figure}

The detector proposed uses the existing superconducting magnet at INFN-LNF for the KLOE experiment \cite{kloe} (Fig. \ref{fig:kloe-stt}).
The magnet provides 0.6\,T magnetic field over a 4.3\,m long, 4.8\,m diameter volume. 
The coil is operated at a nominal current of 2902\,A and the stored energy is 14.32\,MJ.
Inside the magnet, close to the coil cryostat is the cylindrical barrel electromagnetic calorimeter.
Two additional end-caps calorimeters ensure hermeticity along the magnet.  
The barrel calorimeter consists of 24 4.3\,m long, 23\,cm thick modules.
Each end-cap consists of 32 vertical modules 0.7 to 3.9\,m long and 23\,cm thick.  
All modules are stacks of about 200 grooved, 0.5\,mm thick lead foils alternating with 200 layers of cladded 1\,mm diameter scintillating fibers 
which provides high light transmission and sub-ns fast timing.
Photo-cathodes are employed to read both ends of each module.
The energy resolution of the calorimeter is $\sigma/E = 5\%\sqrt(E (GeV))$, and 
the time resolution is $54/\sqrt{E(GeV)}$\,ps.

\section{Physics Opportunities}

Rich physics topics are offered by the design of highly-capable high-resolution magnetized detector. 
In particular, the combination of CH$_2$ target provided by the radiator and pure carbon target provided by graphite, offers an unique opportunity of measuring neutrino-hydrogen interactions, which provide tools to measure the LBNF neutrino flux and constrain systematic uncertainties for DUNE oscillation measurement. 

\subsection{Neutrino-Hydrogen Interaction}

\begin{figure}[htb]
\centering
\includegraphics[height=1.5in]{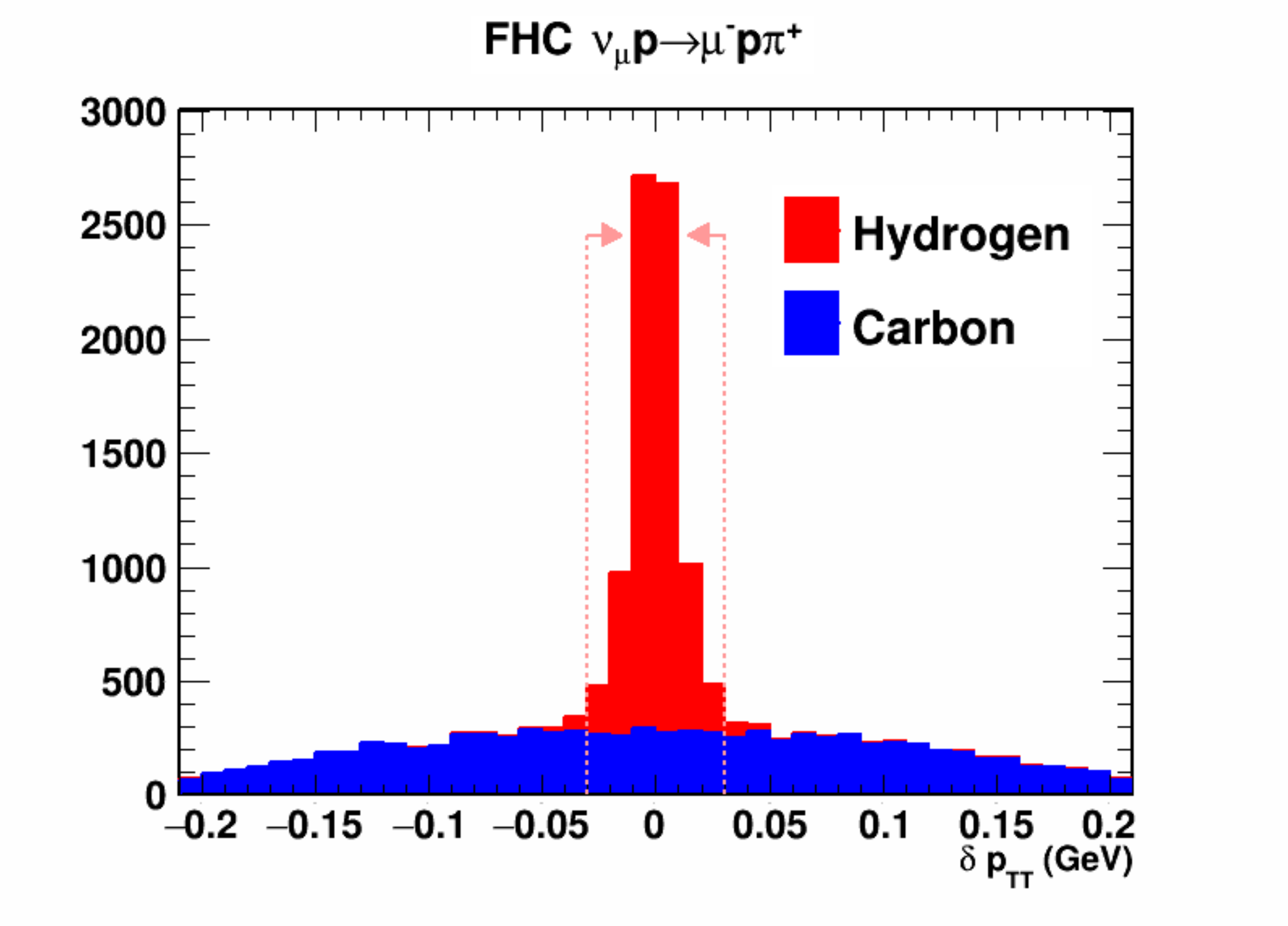}
\includegraphics[height=1.5in]{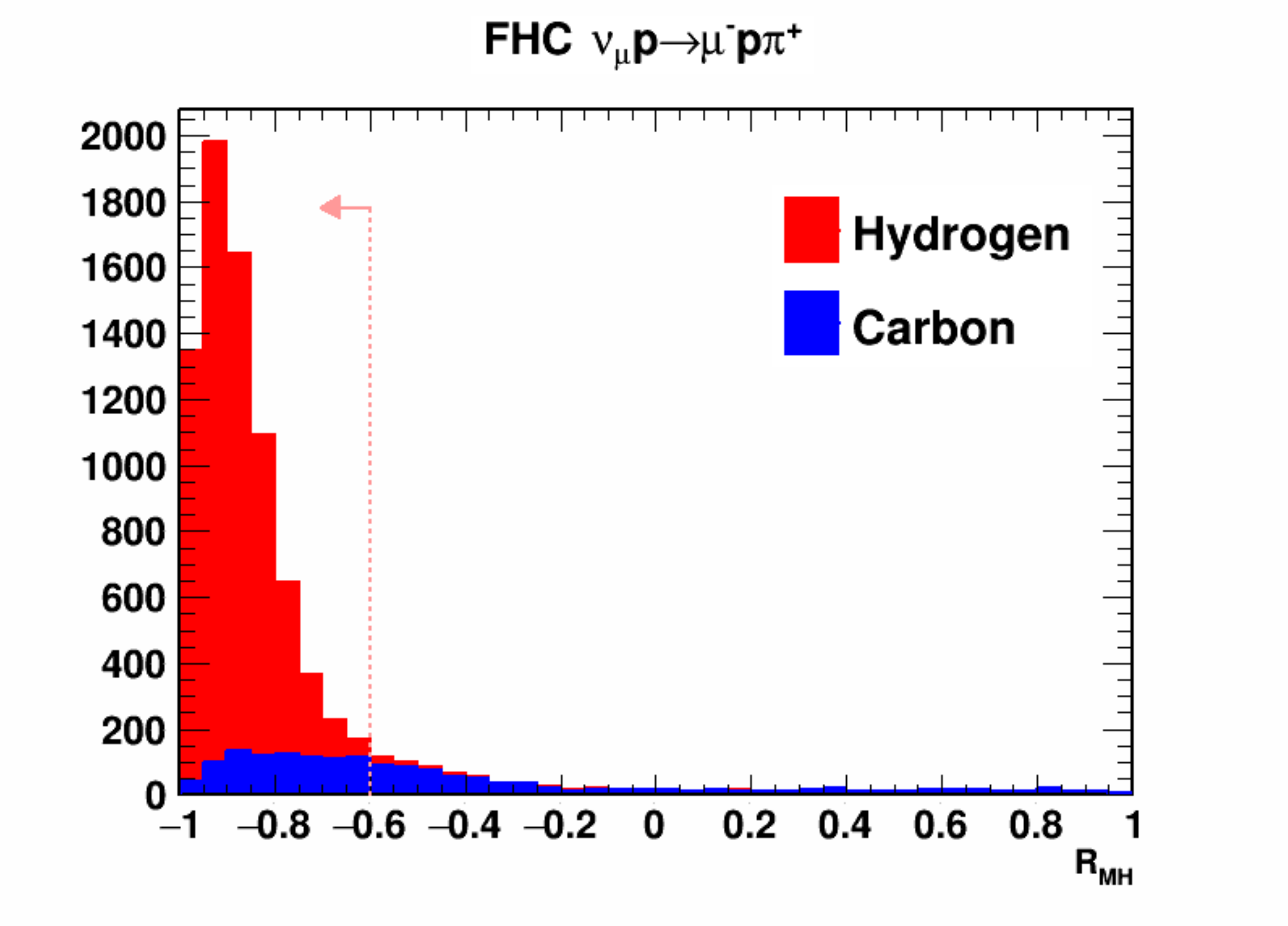}
\includegraphics[height=1.5in]{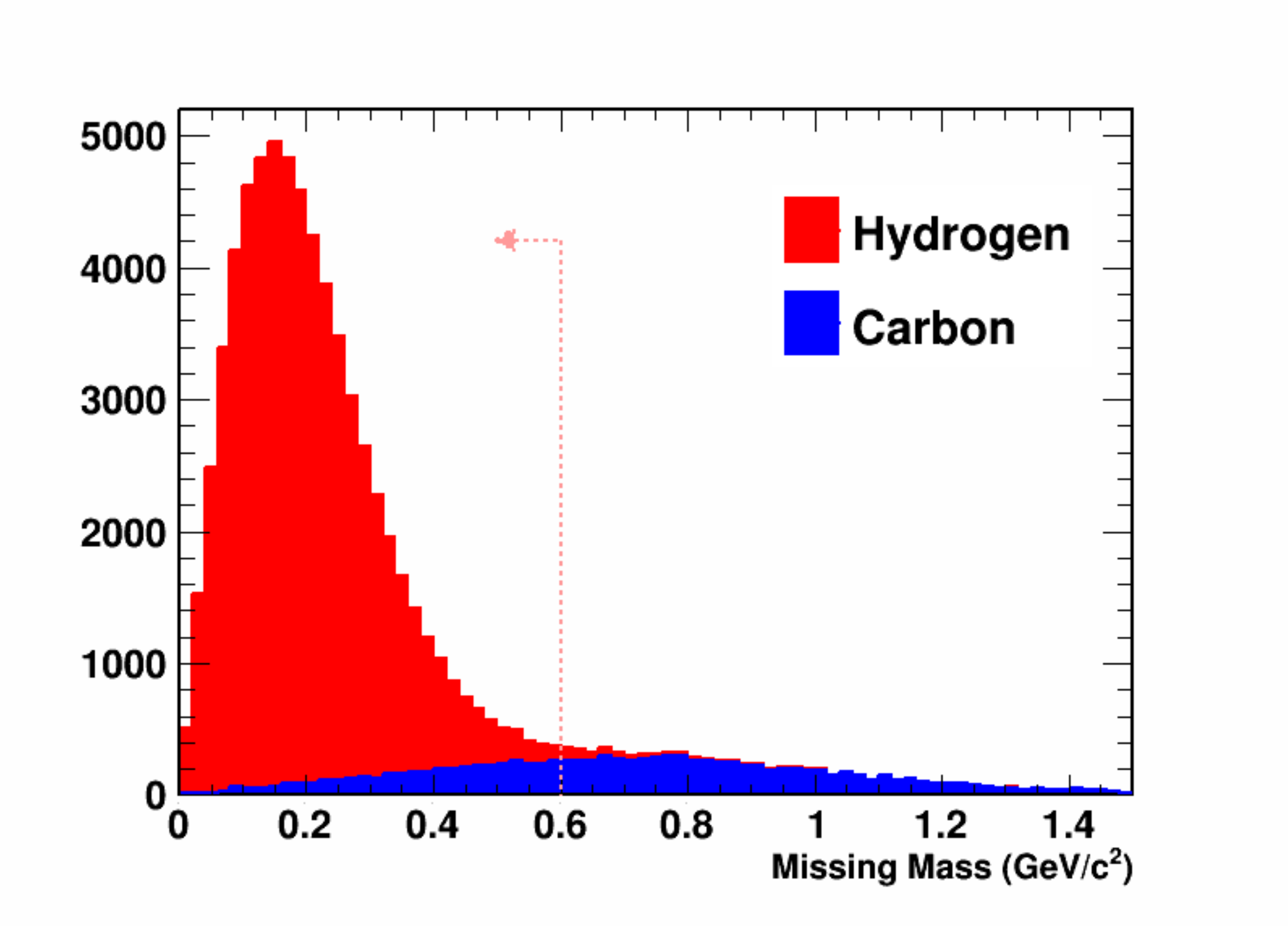}
\caption{Example of neutrino-hydrogen interactions ($\nu_{\mu}p\rightarrow\mu^-p\pi^+$) in STT. Simulated neutrino-hydrogen interactions are shown in red, and neutrino-carbon interactions are in blue. Three kinematics variables are shown:  momentum imbalance on the transverse direction to both incoming neutrino and outgoing muon ($\delta p_{TT}$), $r_{MH} = (p^T_{M} - p^T_{Had})/(p^T_{M} + p^T_{Had})$ where $p^T_{M}$ is the missing momentum traverse to incoming neutrino direction, and $p^T_{Had}$ is the total hadronic momentum traverse to incoming neutrino direction; reconstructed neutrino mass using 4-momentums of final state particles (missing mass). The  dash line and arrows showing the cuts used to select neutrino-hydrogen signal regions. The $r_{MH}$ plot has $\vert p_{TT}\vert<0.03$\,GeV applied, and the missing mass plot has both $\vert p_{TT}\vert<0.03$\,GeV and $r_{MH}<-0.6$ cuts applied. }
\label{fig:hsel}
\end{figure}
The designed fiducial mass of 5 tons of CH$_2$ radiator targets provides about 714 kg of hydrogen.
Neutrino-hydrogen interactions are of interest to neutrino interaction physics.
Moreover, it provides a sample free from nuclear effects to reduce systematic uncertainties for long-baseline oscillation measurements. 
The key of selecting neutrino-hydrogen interactions is the fact that without nuclear effects the momentums of the final state particles are balanced in the direction transverse to the incoming neutrino direction.  
Kinematic variables are developed showing separation between $\nu$-H and $\nu$-C interactions. 
Examples of such variables are shown in Fig \ref{fig:hsel}.
Combined they are able to select a high-purity $\nu$-H sample with high efficiency and greatly reduced $\nu$-C background.
The remaining $\nu$-C background is then measured by the interactions on the graphite target and subtracted from data. 
The subtraction is totally data-driven and model-independent. 
More details of $\nu$-H selection can be find in \cite{duyang_hydrogen}.

\subsection{Neutrino Flux Measurement}

\begin{figure}[htb]
\centering
\includegraphics[height=2.2in]{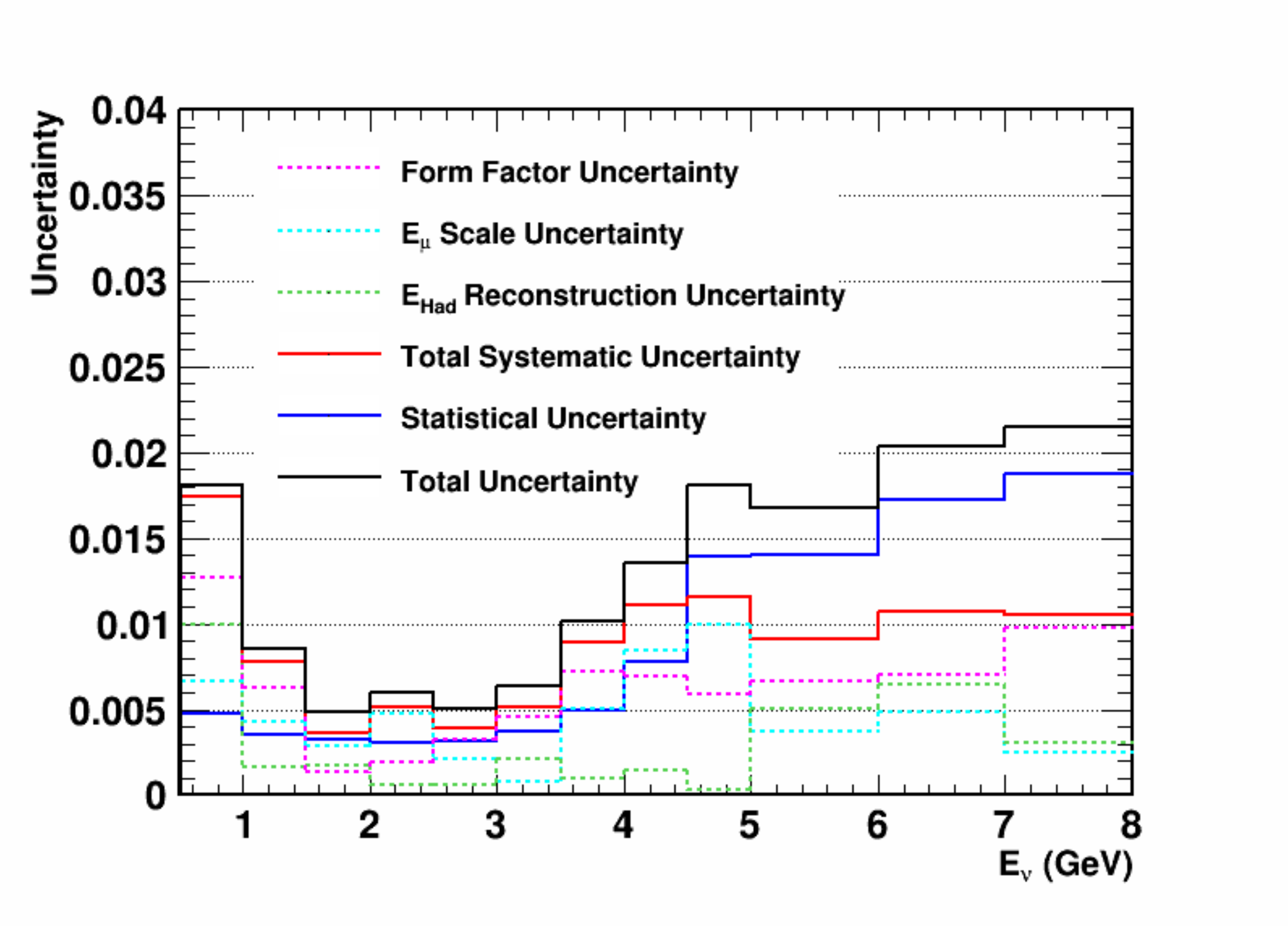}
\includegraphics[height=2.2in]{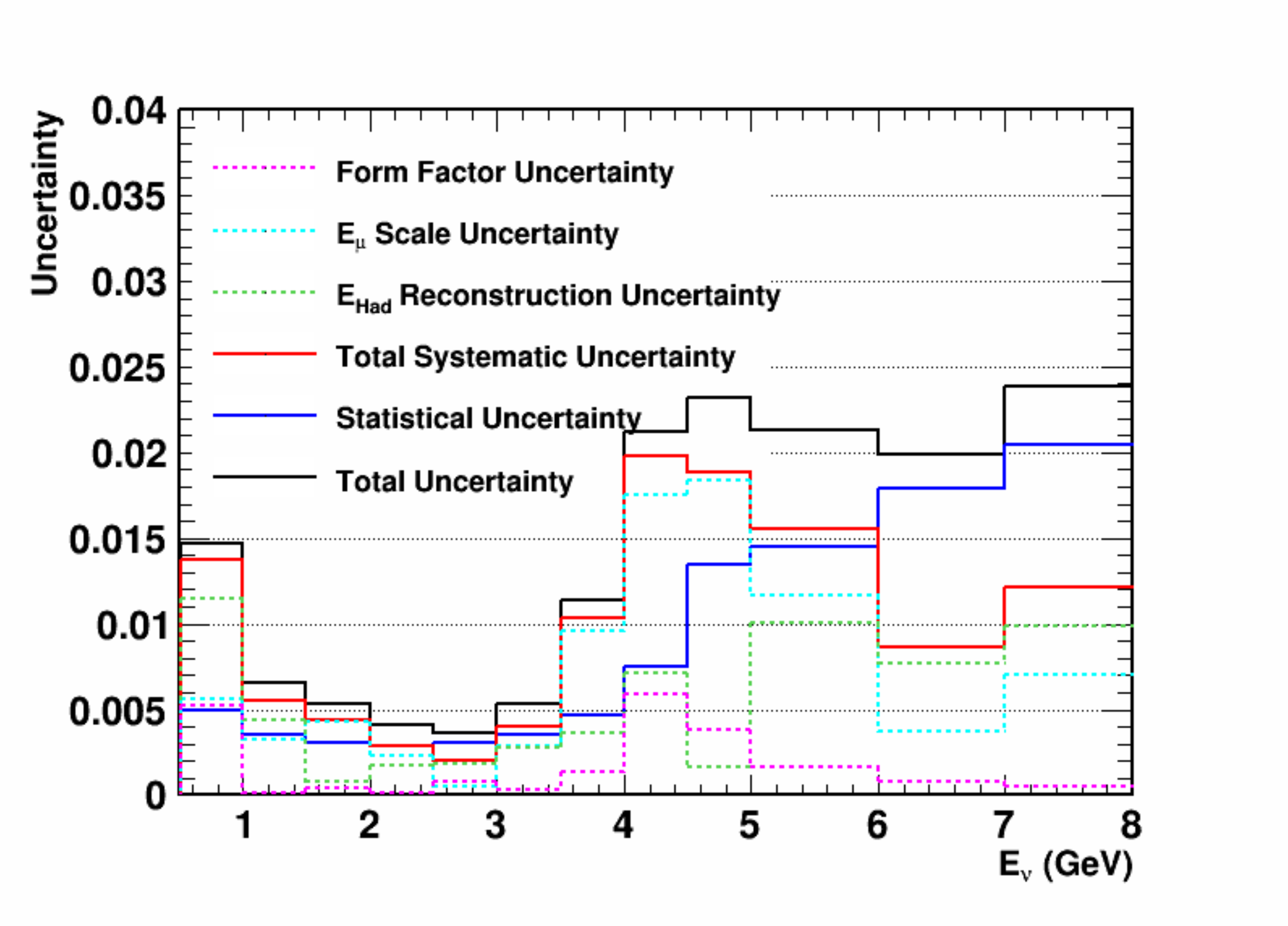}
\caption{Flux shape uncertainties expected from neutrino-hydrogen low-$\nu$ measurements. Two possible channels are shown: $\nu_{\mu}p\rightarrow\nu^-p\pi^+$ in neutrino mode (left), and $\nu_{\mu}n\rightarrow\mu^-p$ in antineutrino mode (right). More details about the flux measurement method and uncertainty can be found in \cite{duyang_flux}}.
\label{fig:flux}
\end{figure}

Knowledge of neutrino flux is critical to neutrino oscillation and cross-section measurements.
A commonly used method in measuring the flux is the low-$\nu$ method, where $\nu = E_{\nu} - E_{l}$ is the energy transfer from the leptons to hadrons. 
At $\nu\ll E_{\nu}$, the cross section of neutrino interaction is flat as function of neutrino energy. 
A measurement of the low-$\nu$ neutrino energy spectrum therefore is a measurement of the neutrino flux shape.
The precision of traditional low-$\nu$ measurement using neutrino-nucleus interactions however is limited by uncertainties from nuclear effects. 
The capability of $\nu$-H measurement in STT has the potential to greatly reduce the uncertainty of low-$\nu$ flux measurement.
It would benefit LBNF physics programs including long-baseline oscillation measurement at DUNE with reduced systemtaic uncertainty. 
An example of such measurement is shown in Fig \ref{fig:flux} using the proposed LBNF flux at the DUNE near site.  
More details of low-$\nu$ flux measurement on hydrogen in STT can be find in \cite{duyang_flux}.

\subsection{Nuclear Effects}

Neutrino oscillations are measured as function of neutrino energy.
The reconstruction of neutrino energy is vulnerable to nuclear effects, which could produce low-energy charged particles below detector threshold, or neutrons difficult to detect.  
This is especially important for DUNE which uses argon target with large nuclear smearing. 
STT is able to perform measurement of neutrino-hydrogen interactions which is free from nuclear effects. 
By comparing with neutrino interactions on nuclear targets, including argon target in STT with similar detector response, the neutrino-hydrogen measurement offers an unique opportunity to measure nuclear effects and constrain neutrino energy scale. 

\subsection{Others}
The low-threshold, high-resolution measurements in STT also provide excellent opportunities to study neutrino-nucleus interactions on various nuclear targets with high statistics. 
In addition, a number of interesting physics topics are available, including electroweak precision measurements, tests of isospin physics and sum rules, measurement of nucleon structure etc.. 

\section{Summary}

We are proposing a low-density, high-resolution, magnetized detector at the near site of the LBNF beamline, including a new straw-tube tracker and reuse of KLOE magnet and electromagnetic calorimeter. 
This combination provides a low-budget solution to enhance the LBNF physics programs. 
It could be part of the DUNE near detector complex, or an stand-alone project. 
Either of the options could benefit the long-baseline neutrino oscillation measurement of DUNE with reduction of systematic uncertainties from neutrino flux and nuclear effects, and offers abundant physics topics by itself, 



\end{document}